# Tunable double notch filter on thin-film lithium niobate platform

SONGYAN HOU,[1,2,4,*] HAO HU,[3,4] ZHIHONG LIU,[1,2] WEICHUAN XING,[1,2] JINCHENG ZHANG,[1,2,*] YUE HAO[1]

[1] *Guangzhou Institute of Technology, Xidian University, Guangzhou, 510555, China*
[2] *State Key Laboratory of Wide Bandgap Semiconductor Devices and Integrated Technology, School of Microelectronics, Xidian University, Xi'an 710071, China*
[3] *Key Laboratory of Radar Imaging and Microwave Photonics Ministry of Education, College of Electronic and Information Engineering, Nanjing University of Aeronautics and Astronautics, Nanjing 211106, China*
[4] *These authors contribute equally*
*\*Songyan Hou  housy1@outlook.com; Jincheng Zhang_jchzhang@xidian.edu.cn*

**Tunable optical filter at the chip scale plays a crucial role in fulfilling the need for the reconfigurability in channel routing, optical switching, and wavelength division multiplexing systems. In this letter, we propose a tunable double notch filter on thin-film lithium niobate using dual micro-ring architecture. This unique integrated filter is essential for complex photonic integrated circuits, along with multiple channels and various frequency spacing. With only one loaded voltage, the device demonstrates a wide frequency spacing tunability from 16.1 GHz to 89.9 GHz by reversely tunning the resonances of the two micro-rings while the center wavelength between the two resonances remains unaltered. Moreover, by utilizing the pronounced electro-optic properties of lithium niobate, associated with the tight light confinement nanophotonic waveguides, the device demonstrates a spacing tunability of 0.82 GHz/V and a contrast of 10~16 dB. In addition, the device has an ultracompact footprint of 0.0248 mm$^2$.**

Integrated optical filter is one of fundamental building blocks in many optical applications, notably in the context of wavelength division multiplexing (WDM) networks[1, 2]. To meet the flexibility requirements of the wavelength selection in the forthcoming highly functional optical circuits, the optical filter featuring continuous tunability has attracted increasing attention. To realize the tunable optical filter, widely adopted approaches are implemented based on silicon photonics. For example, arrayed waveguide grating[3] and cascaded micro-ring filter[4] on silicon platform have been demonstrated a wide tunability. However, the lack of intrinsic electro-optic effect in silicon photonics makes the optical filter suffer from extra optical transmission losses and high power consumption when the functionality is achieved by the means of carriers doping and thermal tunning, prompting the exploration of alternative materials with robust Pockels electro-optic coefficient, such as lithium niobate (LN)[5-10].

Recent developments in thin-film LN nanophotonics have enabled the integration of filters and resonators into a range of photonic architectures, including photonic crystal nanobeams[11], Bragg grating[12-14] and racetrack resonators[15, 16]. However, the thin-film LN optical filters based on add-drop racetrack resonator by thermally tunning resonance and Bragg grating resonator have several drawbacks such as high optical loss, multiple applied voltages, and high power consumption. Compared with them, the optical filter based on microrings exhibits significant potential in tunable filters owing to their inherent characteristics, such as compact footprint, low optical losses, and wavelength tunability. As such, the thin-film LN optical filters based on microrings are supposed to have the superior performance, that combining the advantages of thin-film LN in excellent electro-optic properties and compact structure in reconfigurability[17, 18].

While the microring based filter have been widely reported previously[19, 20], such single ring based filters are not entirely conducive to precise channel filtration of correlated photons, where two or more ultra-narrowband channels separation is critically needed for symmetric filtration of correlated signal and idler waves.

In this Letter, a tunable double notch filter is proposed and experimentally demonstrated on thin-film LN. The design consists of two cascaded microring resonators, whose electric filed are revised by one external voltage. The frequency spacing of the filter is tunable by reversely tuning the resonances of the two microrings. A wide separation tunability range from 16.1 GHz to 89.9 GHz is demonstrated, and the separation can be increased to free spectral range (FSR) of 423 GHz when further increase the voltage. With an ultracompact footprint of 0.0248 mm$^2$, the device provides a frequency separation tunability of 0.82 GHz/V with a contrast of 10~16 dB.

Fig.1 shows the photonic architecture of our designed tunable double notch filter, made of two cascaded microring resonators and a waveguide. As shown in Fig. 1(a), the light is guided into the optical waveguide and coupled with two microrings but experienced with reversed phase shifter in the two microrings. The directions of electric field vector are reversed with a single loaded voltage, seeing the arrows in the cross sections of two microring shown in Fig. 1 (b) and Fig. 1(c). The opposite direction of electric field vectors results in the positive and negative refractive index change in the two microring, and thus positive and negative phase shifters, i.e., $\Delta\phi$ and $-\Delta\phi$.

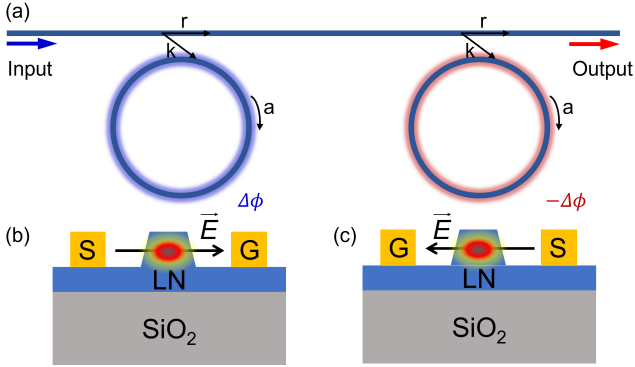

Fig. 1. (a) Architecture design of the double notch filter based on two cascaded microrings. *r* represents the self-coupling coefficient at the coupler; *k* is the cross-coupling coefficient; and *a* represents the single-pass transmission amplitude. (b,c) Cross-section view of the transverse electric mode in thin-film LN waveguide in first (a) and second (b) microrings. The two arrows show the directions of the applied electric field vectors in two microrings.

To simplify the filter model, we assume that the optical transmission coefficients and optical power coupling coefficients are identical in the model. The transmission function T is the result of the product of two microrings, and can be expressed as:

$$T = T_{ring1} \times T_{ring2} = \left(\frac{a^2 - 2ra\cos\phi_1 + r^2}{1 - 2ra\cos\phi_1 + (ra)^2}\right) \times \left(\frac{a^2 - 2ra\cos\phi_2 + r^2}{1 - 2ra\cos\phi_2 + (ra)^2}\right) \quad (1)$$

where r represents the self-coupling coefficient at coupler; k is the cross-coupling coefficient; and a represents the single-pass transmission amplitude. $\phi_1$ and $\phi_2$ are the phase shift after single pass in the microring and are defined as: $\phi = \frac{2\pi nL}{\lambda}$.

To illustrate the working principle of the design, theoretical calculation was performed with *r*=0.899, *a*=0.8998, ring radius R=40 um. Consider that the two microrings are identical and applied with reversed electric fields, then the reflective indices of the two microrings become $n_1 = n - \Delta n$ and $n_2 = n + \Delta n$, respectively (shown in Fig. 2(a)), giving rise to $\phi_1 = \phi - \Delta\phi$ and $\phi_1 = \phi + \Delta\phi$ respectively. Here, the $\Delta n$ and $\Delta\phi$ are the refractive index change and phase shift induced by the reversed electric field, respectively. Then the transmission T is now written as:

$$T = \left(\frac{a^2 - 2ra\cos(\phi - \Delta\phi) + r^2}{1 - 2ra\cos(\phi - \Delta\phi) + (ra)^2}\right) \times \left(\frac{a^2 - 2ra\cos(\phi + \Delta\phi) + r^2}{1 - 2ra\cos(\phi + \Delta\phi) + (ra)^2}\right) \quad (2)$$

Fig. 2(b) shows the calculated transmission responses of the filter as a function of the applied voltages. When various voltage applied, the resonant wavelength of the two microrings shifts to opposite directions, resulting in tunable channel spacing. And the maximum channel spacing of the filter is limited by the FSR.

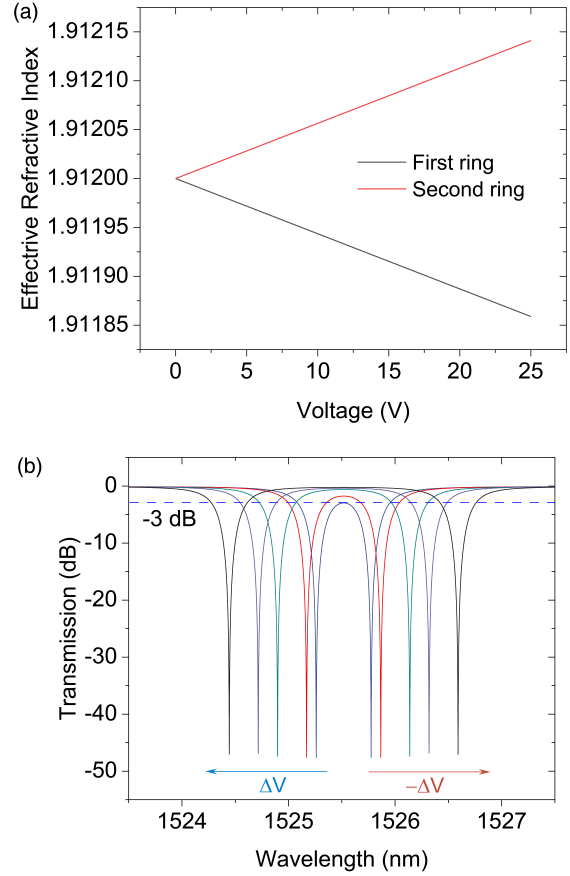

Fig. 2 (a) Calculated effective refractive index of the two microrings as a function of applied voltages. (b) Calculated transmission spectrum with different applied voltages.

The optical filter is fabricated with standard nanofabrication processes[21] on a x-cut 600 nm thin-film LN on silica wafer (Fig. 3(a)). The top width of the LN waveguide is defined as 1.5 um using electron beam lithography (EBL) with negative resist (FOX 16)[18, 22-25]. To optimize the coupling efficiency, the LN waveguide is tapered to 3 um at chip edge (Fig. 3(b)). Using argon based inductively coupled plasma-reactive ion etching (ICP-RIE), the pattern is then transferred to the thin-film LN, leaving 200 nm LN slab. The radius of the on-chip microring is designed as 40 um with electrode gap of 4.5 um (Fig. 3(a)). PMMA/MAA copolymer photoresists are used to define electrodes patterns and 300 nm thick gold electrodes are formed by ebeam evaporation and lift-off processes. A 1.5 um thick silicon dioxide cladding layer is deposited using plasma enhanced chemical vapor deposition. The via windows are formed by maskless photolithography and BOE wet etching. The top metal strips and contact electrode pad are produced by another photolithography, ebeam evaporation and lift-off processes. Fig. 3(c) shows the experimental setup for the optical filter measurement. The transmission spectrum of the filter is measured using a tunable light source and photodetector. The optical mode of light source is controlled by a FPC and then amplified by an EDFA before coupled into the photonic chip. The transmission spectrum is controlled by a voltage controller and the light is coupled in/out by a lensed fiber coupler. PD and OSA are used to monitor the optical power.

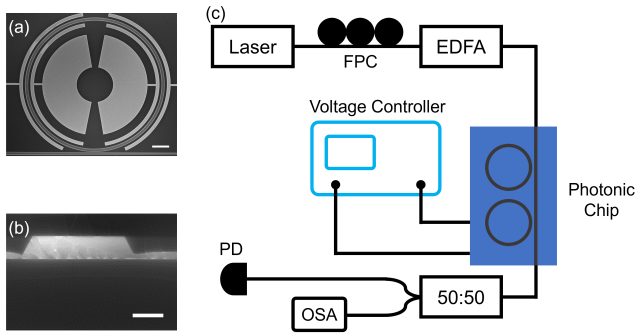

Fig. 3. (a) Scanning electron microscope image of the fabricated ring resonator with electrodes. Scale bar: 10um. (b) Scanning electron microscope image of the cleaved waveguide taper cross section. Scale bar: 1um. (c) Illustration of the experimental setup. FPC: fiber polarization controller; EDFA: erbium-doped fiber amplifier; PD: photodetector; OSA: optical spectrum analyzer.

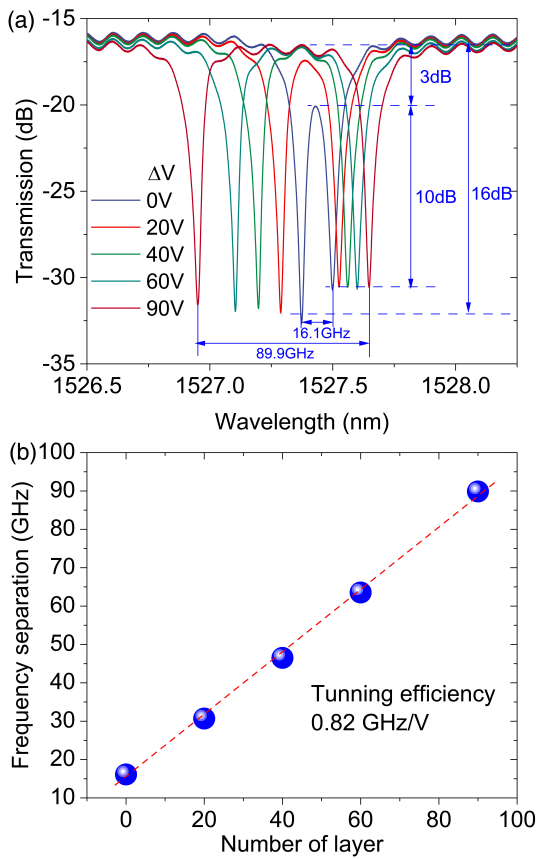

Fig. 4 (a) Transmission spectrum of the filter as a function of applied voltages. (b) Frequency separation of the double notch filter as a function of applied voltages.

Fig. 4(a) shows the measured optical transmission spectrum of the device as a function of applied voltages. The two microrings at static state demonstrate different resonant wavelength and extinction ratio, largely due to the fabrication fluctuations. When the voltage applied, the refractive index of waveguides in two microrings changes reversely, leading to the resonant wavelength of the two microrings shifts into different directions, and results in a peak at the center of the two resonances. The extinction of the resonance is slightly decreasing when the voltage increases. This is due to the transmission spectrum is a product of the two microrings, and the overlap of the two resonance drops when the resonances shift into reverse directions. The 3-dB minimum frequency separation we obtained here is 16.1 GHz (Fig. 4(a)), and the separation is increased to 89.9 GHz when 90 V is applied to the filter while the center wavelength remains unaffected. This separation can be further tuned to FSR bandwidth of 423 GHz if the voltage is continuously increased. The lower drive voltage can be achieved by increasing the effective modulation length at the cost of a reduced FSR, which is the limitation of this device. The device shows a frequency separation tunability of 0.82 GHz/V with a contrast of 10~16 dB. The extinction ratio can be improved by optimizing the waveguide and coupler parameters including waveguide width, coupler type, ring radius, etc[26]. By utilizing the pronounced electro-optic properties of LN, coupled with the tight light confinement nanophotonic waveguides, compact footprint of 0.0248 $mm^2$ is achieved. Further improvement of tunning efficiency can be achieved by using two racetracks with longer modulation length.

In conclusion, we have demonstrated a tunable double notch optical filter based on two cascaded microrings on thin-film LN platform. With only one external voltage, the optical waves inside the two loaded microrings experience reversed external electric field, resulting in different refractive index changes and opposite resonance shifts. The device demonstrates a minimum frequency separation of 16.1 GHz and the spacing increases to 89.9 GHz when 90 V is applied to the filter. The separation can further span to FSR bandwidth of 423 GHz with higher voltage. Utilizing the pronounced electro-optic properties of lithium niobate, coupled with the tight light confinement nanophotonic waveguides, the device demonstrates a spacing tunability of 0.82 GHz/V, while the center wavelength remains unaltered. The optical filter has an extinction ratio of 10~16 dB with an ultracompact footprint of 0.0248 $mm^2$. Owing to large channel separation tunability, ultracompact size and low power consumption, the proposed optical filter on thin-film LN shows great potential for effective channel isolation in the next generation WDM optical networks. Our work not only facilitates the applications of thin-film LN in high-performance optical filters, but also opens up a new opportunity to explore light-matter interactions in the emerging time-varying systems[27].

**Funding.** National Key R&D Program of China (grant no. 2020YFB1807300); Natural Science Basic Research Program of Shaanxi Province (2024JC-YBQN-0682).

**Disclosures.** The authors declare no conflicts of interest.

**Data availability.** Data underlying the results presented in this paper are not publicly available at this time but may be obtained from the authors upon reasonable request.

## References

1. R. Magnusson, and S. Wang, "New principle for optical filters," Applied physics letters **61**, 1022-1024 (1992).
2. M. Uenuma, and T. Motooka, "Temperature-independent silicon waveguide optical filter," Opt. Lett. **34**, 599-601 (2009).
3. D. Liu, H. Wu, and D. Dai, "Silicon multimode waveguide grating filter at 2 μm," Journal of Lightwave Technology **37**, 2217-2222 (2019).
4. T. Dai, A. Shen, G. Wang, Y. Wang, Y. Li, X. Jiang, and J. Yang, "Bandwidth and wavelength tunable optical passband filter based on silicon multiple microring resonators," Opt Lett **41**, 4807-4810 (2016).
5. Z. He, H. Guan, X. Liang, J. Chen, M. Xie, K. Luo, R. An, L. Ma, F. Ma, T. Yang, and H. Lu, "Broadband, Polarization-Sensitive, and Self-Powered High-Performance Photodetection of Hetero-Integrated $MoS_2$ on Lithium Niobate," Research **6**, 0199 (2023).
6. C. Wang, M. Zhang, X. Chen, M. Bertrand, A. Shams-Ansari, S. Chandrasekhar, P. Winzer, and M. Loncar, "Integrated lithium niobate electro-optic modulators operating at CMOS-compatible voltages," Nature **562**, 101-104 (2018).
7. M. He, M. Xu, Y. Ren, J. Jian, Z. Ruan, Y. Xu, S. Gao, S. Sun, X. Wen, L. Zhou, L. Liu, C. Guo, H. Chen, S. Yu, L. Liu, and X. Cai, "High-performance hybrid silicon and lithium niobate Mach–Zehnder modulators for 100 Gbit s1 and beyond," Nature Photonics **13**, 359-364 (2019).
8. Q. Luo, C. Yang, Z. Hao, R. Zhang, R. Ma, D. Zheng, H. Liu, X. Yu, F. Gao, F. Bo, Y. Kong, G. Zhang, and J. Xu, "On-chip erbium–ytterbium-co-doped lithium niobate microdisk laser with an ultralow threshold," Opt. Lett. **48**, 3447-3450 (2023).
9. D. Zhu, L. Shao, M. Yu, R. Cheng, B. Desiatov, C. Xin, Y. Hu, J. Holzgrafe, S. Ghosh, and A. Shams-Ansari, "Integrated photonics on thin-film lithium niobate," Advances in Optics and Photonics **13**, 242-352 (2021).
10. M. Xu, M. He, H. Zhang, J. Jian, Y. Pan, X. Liu, L. Chen, X. Meng, H. Chen, Z. Li, X. Xiao, S. Yu, S. Yu, and X. Cai, "High-performance coherent optical modulators based on thin-film lithium niobate platform," Nature Communications **11**, 3911 (2020).
11. M. Li, J. Ling, Y. He, U. A. Javid, S. Xue, and Q. Lin, "Lithium niobate photonic-crystal electro-optic modulator," Nature Communications **11**, 4123 (2020).
12. A. Prencipe, M. A. Baghban, and K. Gallo, "Tunable Ultranarrowband Grating Filters in Thin-Film Lithium Niobate," ACS Photonics **8**, 2923-2930 (2021).
13. M. R. Escalé, D. Pohl, A. Sergeyev, and R. Grange, "Extreme electro-optic tuning of Bragg mirrors integrated in lithium niobate nanowaveguides," Opt. Lett. **43**, 1515-1518 (2018).
14. K. Abdelsalam, E. Ordouie, M. G. Vazimali, F. A. Juneghani, P. Kumar, G. S. Kanter, and S. Fathpour, "Tunable dual-channel ultra-narrowband Bragg grating filter on thin-film lithium niobate," Opt Lett **46**, 2730-2733 (2021).
15. C. Wang, M. Zhang, X. Chen, M. Bertrand, A. Shams-Ansari, S. Chandrasekhar, P. Winzer, and M. Lončar, "Integrated lithium niobate electro-optic modulators operating at CMOS-compatible voltages," Nature **562**, 101-104 (2018).
16. D. Jia, R. Zhang, C. Yang, Z. Hao, X. Yu, F. Gao, F. Bo, G. Zhang, and J. Xu, "Electrically tuned coupling of lithium niobate microresonators," Opt. Lett. **48**, 2744-2747 (2023).
17. M. Zhang, C. Wang, R. Cheng, A. Shams-Ansari, and M. Lončar, "Monolithic ultra-high-Q lithium niobate microring resonator," Optica **4** (2017).
18. S. Hou, H. Hu, W. Xing, Z. Liu, J. Zhang, and Y. Hao, "Improving Linewidth and Extinction Ratio Performances of Lithium Niobate Ring Modulator Using Ring‐Pair Structure," Advanced Photonics Research **4** (2023).
19. Y. Ding, M. Pu, L. Liu, J. Xu, C. Peucheret, X. Zhang, D. Huang, and H. Ou, "Bandwidth and wavelength-tunable optical bandpass filter based on silicon microring-MZI structure," Optics express **19**, 6462-6470 (2011).
20. R. Grover, T. A. Ibrahim, S. Kanakaraju, L. Lucas, L. C. Calhoun, and P. T. Ho, "A tunable GaInAsP-InP optical microring notch filter," IEEE Photonics Technology Letters **16**, 467-469 (2004).
21. S. Hou, A. Xie, Z. Xie, L. Y. M. Tobing, J. Zhou, L. Tjahjana, J. Yu, C. Hettiarachchi, D. Zhang, C. Dang, E. H. T. Teo, M. D. Birowosuto, and H. Wang, "Concurrent Inhibition and Redistribution of Spontaneous Emission from All Inorganic Perovskite Photonic Crystals," ACS Photonics **6**, 1331-1337 (2019).
22. I. Briggs, S. Hou, C. Cui, and L. Fan, "Simultaneous type-I and type-II phase matching for second-order nonlinearity in integrated lithium niobate waveguide," Opt Express **29**, 26183-26190 (2021).
23. P. K. Chen, I. Briggs, S. Hou, and L. Fan, "Ultra-broadband quadrature squeezing with thin-film lithium niobate nanophotonics," Opt Lett **47**, 1506-1509 (2022).
24. S. Hou, P. Chen, M. Shah, I. Briggs, W. Xing, Z. Liu, and L. Fan, "Programmable Optical Filter in Thin-Film Lithium Niobate with Simultaneous Tunability of Extinction Ratio and Wavelength," ACS Photonics (2023).
25. M. Shah, I. Briggs, P. K. Chen, S. Hou, and L. Fan, "Visible-telecom tunable dual-band optical isolator based on dynamic modulation in thin-film lithium niobate," Opt Lett **48**, 1978-1981 (2023).
26. D. T. Spencer, J. F. Bauters, M. J. R. Heck, and J. E. Bowers, "Integrated waveguide coupled $Si_3N_4$ resonators in the ultrahigh-Q regime," Optica **1** (2014).
27. Y. Yu, H. Hu, L. Zou, Q. Yang, X. Lin, Z. Li, L. Gao, and D. Gao, "Antireflection Spatiotemporal Metamaterials," Laser & Photonics Reviews **n/a**, 2300130.